\documentclass[prb,aps,twocolumn]{revtex4}
\usepackage{amsmath}
\usepackage{amssymb}
\usepackage {graphicx}
\usepackage{color}
\def\blue{\color{blue}}

\usepackage{hyperref}
\hypersetup{colorlinks=true,linkcolor=blue,anchorcolor=blue,citecolor=blue,filecolor=blue,urlcolor=blue,bookmarksnumbered=true,pdfview=FitB}
\usepackage{comment}

\newcommand{\beq}{\begin{equation}}
\newcommand{\eeq}{\end{equation}}
\newcommand{\bea}{\begin{eqnarray}}
\newcommand{\eea}{\end{eqnarray}}
\newcommand{\e}{\varepsilon}
\newcommand{\bk}{{\vec k}}

\newcommand{\bq}{{\vec q}}
\newcommand{\bB}{{\vec B}}
\newcommand{\bv}{{\vec v}}
\newcommand{\bes}{\begin{subequations}}
\newcommand{\ees}{\end{subequations}}

\newcommand{\nn}{\nonumber}

\begin{document}

\title{Raman scattering in a two-dimensional Fermi liquid with spin-orbit coupling}

\author{Saurabh Maiti and Dmitrii L. Maslov}
\affiliation{Department of Physics, University of Florida, Gainesville,
FL 32611}

\date{\today}
\begin{abstract}
We present a microscopic theory of Raman scattering in a
two-dimensional Fermi liquid (FL) with Rashba and Dresselhaus
types of spin-orbit coupling, and subject to an in-plane magnetic
field ($\bB$). In the long-wavelength limit, the Raman spectrum
probes the collective modes of such a FL: the chiral spin waves.
The characteristic features of these modes are a linear-in-$q$
term in the dispersion and the dependence of the mode frequency on
the directions of both $\bq$ and $\bB$. All of these features have
been observed in recent Raman experiments on Cd$_{1-x}$Mn$_x$Te
quantum wells.
\end{abstract}
\maketitle

\section{Introduction}
Raman scattering
is an inelastic light scattering process that
allows to study dynamics of elementary excitations in solids
both in space and time and to probe both single-particle and
collective properties of electron systems. It helps to understand
a variety of phenomena: from pairing mechanisms in
high-temperature superconductors\cite{intro1} to {properties of}
spin waves in spintronic devices.\cite{intro2} The latter requires probing dynamics of electron spins, which can be done if the incident and scattered light are polarized perpendicular to each other
(cross-polarized geometry).\cite{raman_book}

Non-trivial spin dynamics is encountered in systems with
spontaneous magnetic order, or in an external magnetic field, or
else in the presence of spin-orbit coupling (SOC). In this
paper, we develop a microscopic theory of Raman scattering in a
two-dimensional (2D) Fermi liquid (FL) in the presence of an
in-plane magnetic field and both Rashba \cite{Rashba} and
Dresselhaus \cite{Dresselhaus} types of SOC. Breaking spin
conservation leads to a substantial modification of the Raman
spectrum already at the non-interacting level. \cite{Magarill}
Unlike in the $SU(2)$-invariant case, the cross-section of Raman
scattering cannot be expressed solely in terms of charge and spin
susceptibilities. As a result, it is not {\em a priori} clear how
single-particle and collective spin-excitations in systems
with SOC {manifest themselves in a} Raman scattering process. We show here that the scattering cross-section, in the cross-polarized
geometry and in the long-wavelength limit, is parameterized by particular components of the spin susceptibility tensor. These components contain poles at frequencies that correspond to {a new type of collective modes--chiral spin waves (CSW)\cite{Shekhter,Ali1,Zhang,SM1,SM_damping,SM_ESR,kumar:2017}--}arising from an interplay between electron-electron interaction ({\em eei}) and {SOC}.

\begin{figure*}[htp]
$\begin{array}{ccc}
\includegraphics[width=0.63\columnwidth,height=1.72in]{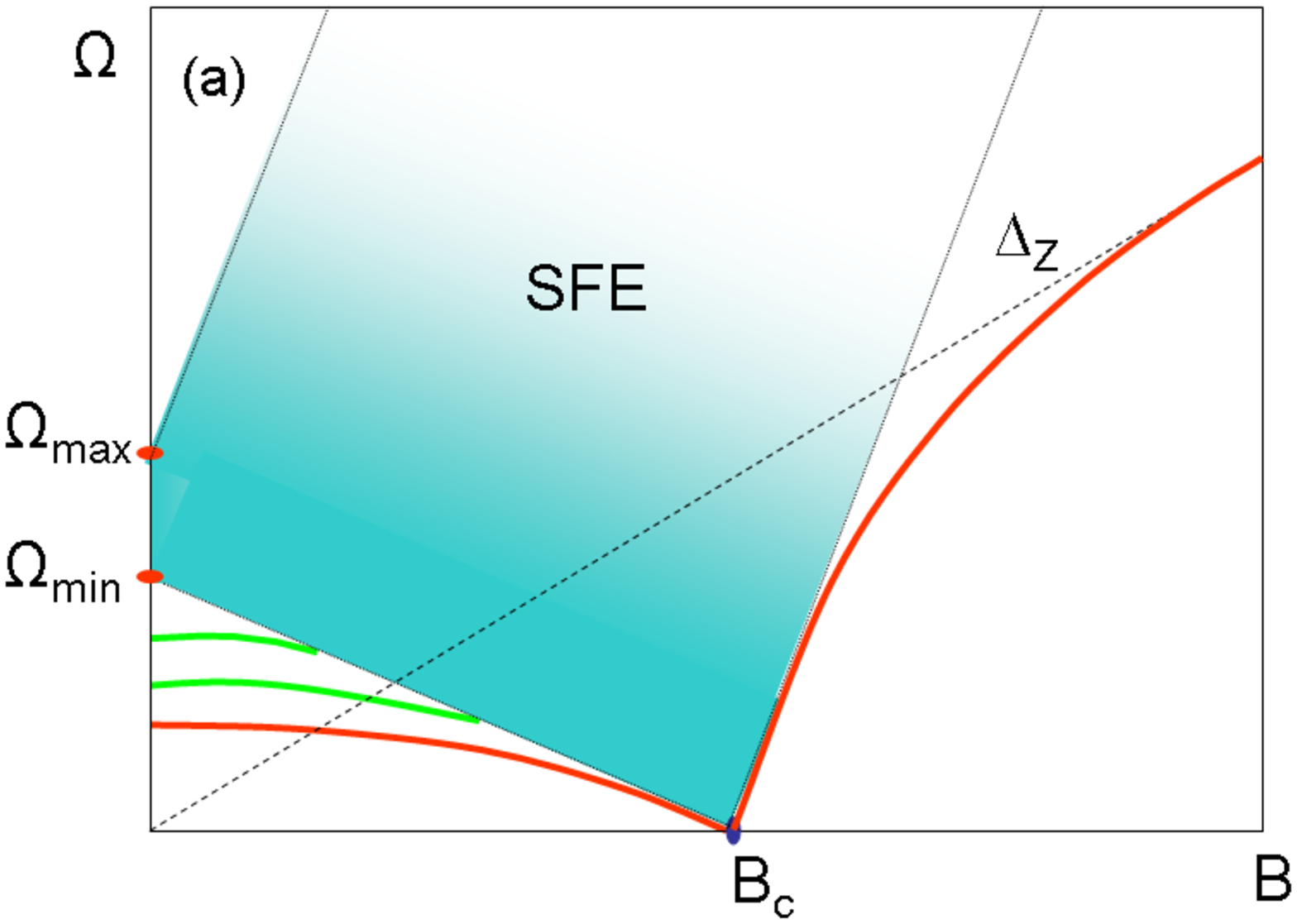}&
\includegraphics[width=0.7\columnwidth]{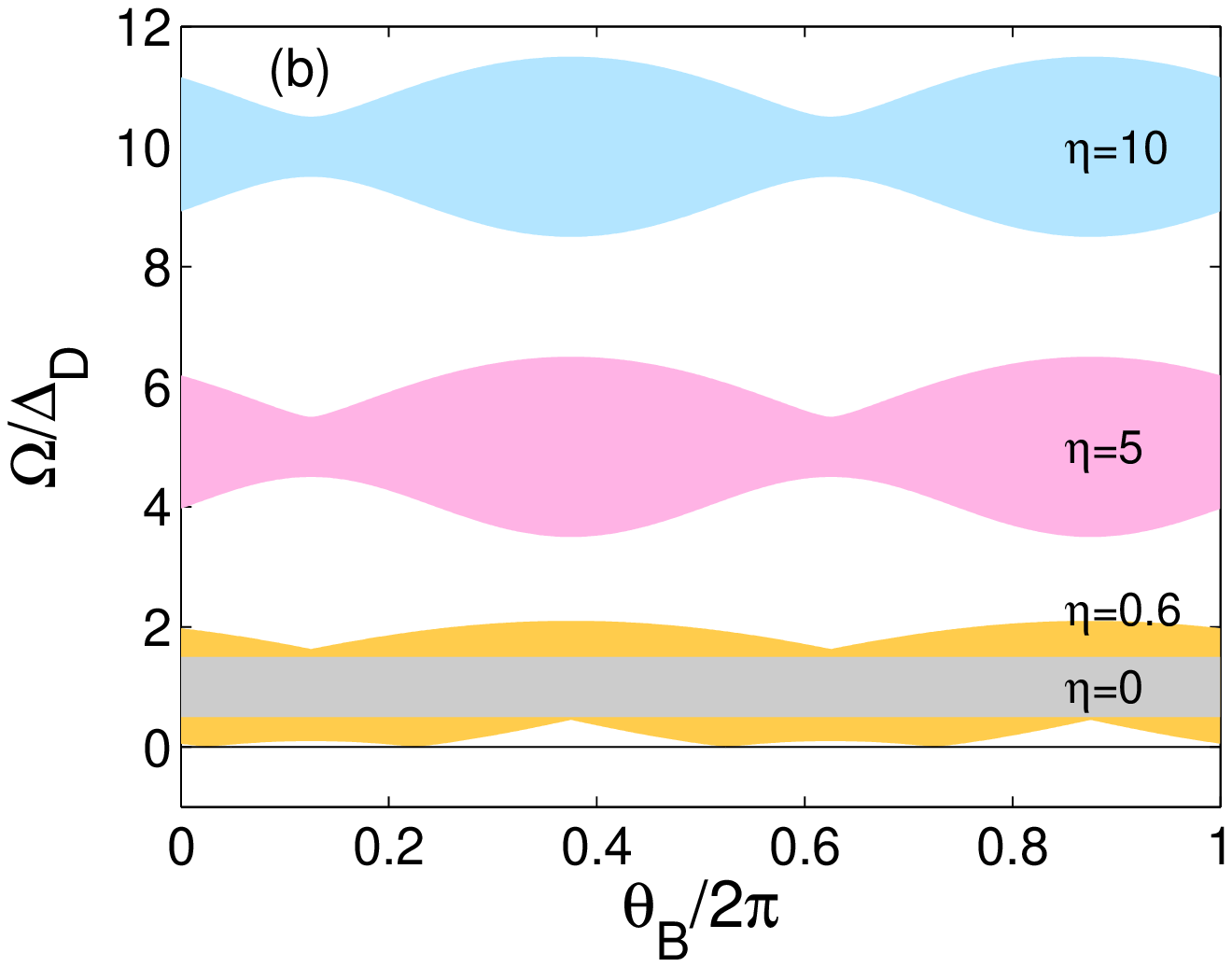}&
\includegraphics[width=0.7\columnwidth]{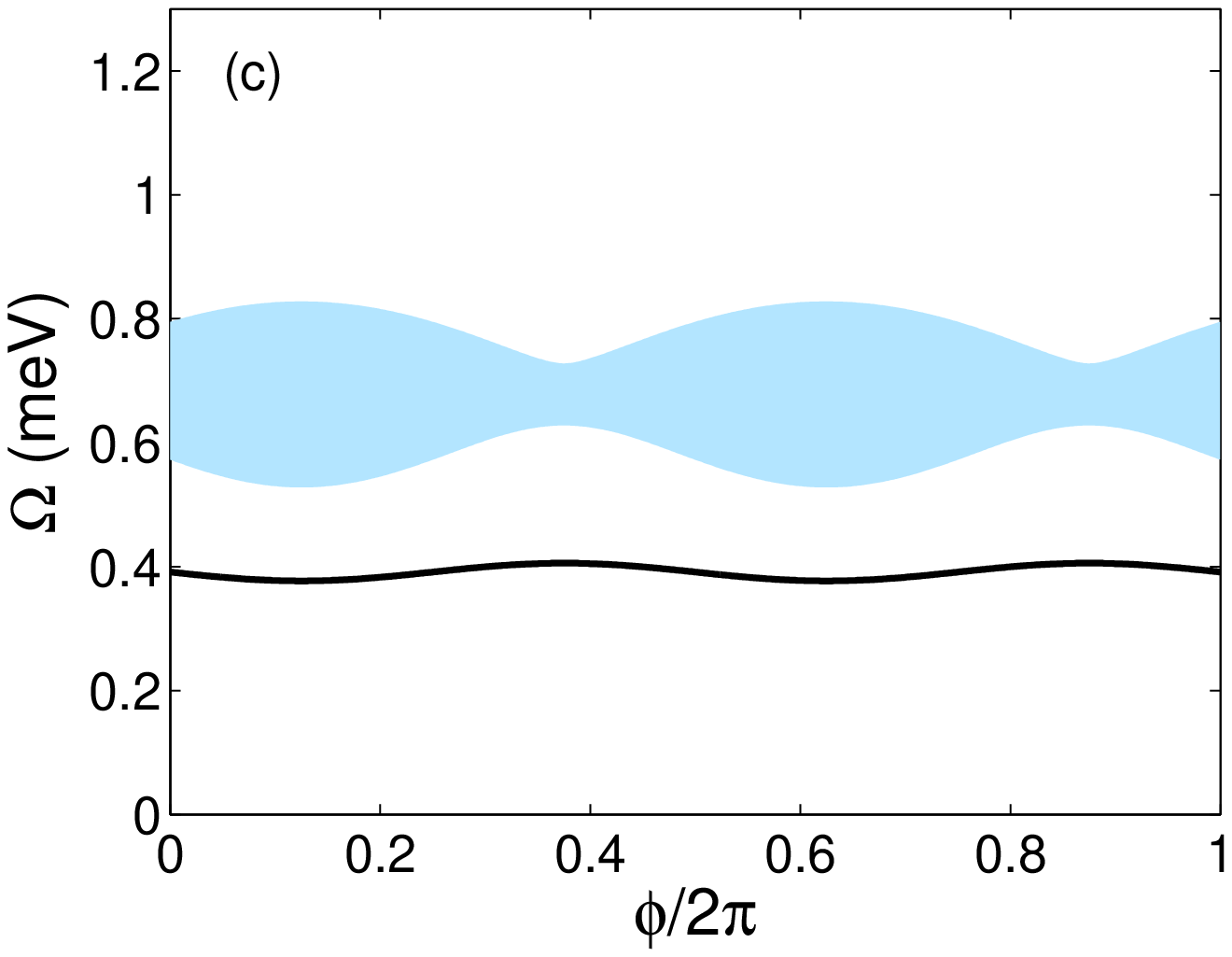}
\end{array}$
\caption{ \label{fig:1}Color online: a) Schematically: excitation
spectrum (at $q=0$) of a 2D Fermi liquid with spin-orbit coupling
and subject to an in-plane magnetic field $\bB$. Shaded region:
continuum of spin-flip excitations; lines: chiral spin waves. b)
Continuum as a function of the angle  $\theta_\bB$ between $\bB$
and the (100) direction.  $\Delta_R/\Delta_D=0.5$,
$\eta=\Delta_Z/\Delta_D$, where $\Delta_{D/R/Z}$ is
Dresselhaus/Rashba/Zeeman splitting.  c) Continuum (shaded) and
collective-mode frequency at $q=0$ (line) as a function of
$\phi=\theta_\bB-\pi/2$ for a  Cd$_{1-x}$Mn$_{x}$Te quantum well.}
\end{figure*}

{In the absence of SOC, the dispersion of a collective mode in the
spin channel must start with a $q^2$ term by analyticity.}
Recently, a dispersing peak was observed in resonant Raman
scattering on magnetically doped CdTe quantum wells in the
presence of an in-plane magnetic field.\cite{BBX1,BBX2,BBX3} The
unusual features observed in these experiments  were a
linear-in-$q$ term in the dispersion and a $\pi$-periodic
modulation of the spectrum as the magnetic field is rotated in the
plane of 2D electron gas (2DEG). {We argue that} the observed peak
{is actually} one of the long sought after CSW
(in the presence of a field). The CSWs are collective oscillations
of the magnetization ($\vec M$) that exist even in the absence of
the magnetic field. In zero field, there are three such modes,
\cite{Shekhter,Zhang,Ali1,SM1} which are massive, i.e.,  their
frequencies are finite at $q\to 0$, and disperse with $q$ on a
characteristic scale set by spin-orbit splitting. The modes are
linearly polarized, with $\vec M$ being in the 2DEG plane for two
of the modes and along the normal for the third one. If an
in-plane magnetic field ($\bB$) is applied, the mode with $\vec M
|| \vec B$ remains linearly polarized while the other two modes
with $\vec M\perp\vec B$ become elliptically
polarized.\cite{SM_ESR} Figure~\ref{fig:1}a depicts the evolution
of the excitation spectrum with $B$ at $q=0$.   As the field
increases, two, out of the three, modes run into the continuum of
spin-flip excitations (SFE). The third mode merges with the
continuum at $B=B_c$, when the spin-split Fermi surfaces (FSs)
become degenerate, and re-emerges to the right of this point. As
the field is increased further, this mode transforms  gradually
into the spin wave [Silin-Leggett (SL) mode] of a partially
polarized FL.\cite{silin:1958,leggett:1970,statphysII} In the
{experimental setup of Refs}.~\onlinecite{BBX1,BBX2,BBX3}, the
effective Zeeman energy is larger than both Rashba and Dresselhaus
splittings which, according to Fig.~\ref{fig:1}{\em a}, allows us
to focus on the case of a single CSW adiabatically connected to
the SL mode. We show that, at small $q$, the dispersion of this
mode can be written as \bea \Omega(\bq,\bB)=
\Omega_0(\theta_\bB)+w(\theta_\bq,\theta_\bB)q+Aq^2, \label{res1}
\eea where $\theta_{\vec x}$ is the azimuthal angle of $\vec x$.
{We show, first by a symmetry argument and then by an explicit
calculation}, that {the} Rashba and Dresselhaus {types of} SOC
contribute $\sin(\theta_\bq-\theta_\bB)$ and
$\cos(\theta_\bq+\theta_\bB)$ terms to $w(\theta_\bq,\theta_\bB)$,
correspondingly, while the mass term, $\Omega_0(\theta_\bB)$ and
the boundaries of the SFE continuum are $\pi$-periodic function of
$\theta_\bB$. Our theory describes quantitatively the results of
Refs.~\onlinecite{BBX1,BBX2,BBX3}, where a single Raman peak
was observed,
and predicts
that up to three peaks  can be observed at lower
magnetic fields in systems with higher mobility and/or stronger SOC.

The rest of the paper is organized as follows. In Sec.
\ref{sec:raman}, we discuss the general theory of Raman response
in a Fermi-liquid with SOC. In Sec.~\ref{sec:symm}, we present
symmetry arguments for the form of the CSW dispersion in the
presence of a magnetic field. In Sec. \ref{sec:micro}, we present
results of an explicit calculation for the CSW dispersion. In
Sec.~\ref{sec:exp} we compare our theory to the experimental
results of Refs.~\onlinecite{BBX1,BBX2,BBX3}. In Sec.
\ref{sec:con}, we summarize our main results. Details of the
calculations are given in Appendix.

\section{General formalism} \label{sec:raman} 
\begin{figure}[htp]
\includegraphics[width=0.9\columnwidth]{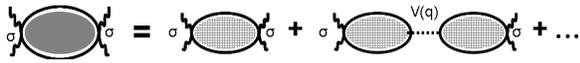}
\caption{ \label{fig:two} Diagrams for the differential
cross-section of Raman scattering in cross-polarized geometry.
Shaded bubbles denote vertex corrections due to the exchange part
of {\em eei}, calculated in the ladder approximation. $V(q)$ is
the bare Coulomb potential.}
\end{figure}

In general, Raman scattering
probes both charge and spin excitations. However, if the
polarizations of the incident ($\vec{\rm e}_{\rm I}$) and
scattered ($\vec{\rm e}_{\rm S}$) light are perpendicular to each
other, the Raman vertex couples directly to the electron
spin.\cite{raman_book,Magarill} The Feynman diagrams for the
differential scattering cross-section in such a geometry are shown
in Fig.~\ref{fig:two}. The first diagram on the right-hand side in the top row
 is a pure spin part, while the
rest is a spin-charge part, which is non-zero due to SOC, and
is represented by a sum of bubbles connected by the Coulomb
interaction (dashed line).

Shading of inner parts of the bubbles in Fig. \ref{fig:two}
denotes renormalizations due to the exchange part  of {\em eei}.
If this renormalization is neglected, the differential (per unit
energy and per unit solid angle) scattering cross-section is
reduced, upon projection onto the conduction band, to the result
of Ref.~\onlinecite{Magarill}:
\begin{subequations}
\bea \frac{d^2{\mathcal A}}{d\Omega d{\mathcal
O}}&\propto&\sum_{\mu\mu'}|\gamma_{\mu\mu'}|^2{\rm
Im}L^0_{\mu\mu'}- \frac{2\pi e^2}{q}{\rm Im}\frac{Z\bar
Z}{\epsilon(q,\Omega)}
,\label{eq:Raman1}\\
\gamma_{\mu\mu'}&=&i\langle\vec n\cdot\hat{\vec\sigma} e^{i\bq\cdot\vec r}\rangle_{\mu\mu'},\\
L^0_{\mu\mu'}&=&
\frac{f(\e_{\mu})-f(\e_{\mu'})}{\Omega+\e_{\mu}-\e_{\mu'}+i\delta},\\
Z&=&\sum_{\mu\mu'}\gamma_{\mu\mu'}
\langle e^{-i\bq\cdot\vec r}\rangle_{\mu\mu'}L^0_{\mu\mu'};\\
\bar Z&=&\sum_{\mu\mu'}\gamma^*_{\mu\mu'}
\langle e^{i\bq\cdot\vec r}\rangle_{\mu\mu'}L^0_{\mu\mu'}\\
\epsilon(q,\Omega)&=&\epsilon+\frac{2\pi
e^2}{q}\sum_{\mu\mu'}|\langle e^{i\bq.\vec
r}\rangle_{\mu\mu'}|^2L^0_{\mu\mu'}. \eea
\end{subequations}
Here, $\epsilon$ is the background dielectric constant, $\mu$ and
$\mu'$ refer to the quantum numbers of electrons which include the
momentum and spin/chirality: $\mu=\{\bk,\nu\}$,
$\mu'=\{\bk',\nu'\}$ with $\nu,\nu'=\pm 1$, $\e_{\mu}$ is the
energy of a state with quantum number $\mu$, $f(\e)$ is the Fermi
function, $\langle X\rangle_{\mu\mu'}\equiv\int {d\vec
r}\psi_{\mu}^{\dagger}(\vec r)X(\vec r)\psi_{\mu'}(\vec r)$ is the
matrix element of $X(\vec r)$ between states $\mu$ and $\mu'$, and
$\vec n = \vec e_{\rm I}\times\vec e_{\rm S}$. The matrix elements
are computed with respect to the eigenvectors of the
non-interacting Hamiltonian for a $(001)$ quantum well with Rashba
and Dresselhaus SOC and in the presence of an in-plane magnetic
field. In what follows, we will be assuming that the energy scales
associated with the magnetic field and SOC are much smaller than
the Fermi energy. In this case, it suffices to use  a low-energy
version of the Hamiltonian \beq\label{eq:1}
\mathcal{H}_{\bk}=v_F(k-k_F)\hat\sigma_0 +
\sum_{i=1,2}s_i\hat\sigma_i, \eeq where $\hat\sigma_0$ is the
identity matrix, $v_F$ and $k_F$ are the Fermi velocity and
momentum in the absence of SOC and magnetic field,
correspondingly, and vector ${\vec s}$ parametrizes the effects of
SOC and magnetic field: \bea\label{eq:temp}
s_1&=&\frac{1}{2}\left(\Delta_R\sin\theta_\bk+\Delta_D\cos\theta_\bk+\Delta_Z\cos\theta_\bB\right),\nn\\
s_2&=&-\frac{1}{2}\left(\Delta_R\cos\theta_\bk+\Delta_D\sin\theta_\bk-\Delta_Z\sin\theta_\bB\right),
\eea where $\Delta_R$, $\Delta_D$, and $\Delta_Z$ are the Rashba,
Dresselhaus, and Zeeman energy splittings, correspondingly. The
$x_1$ axis is chosen along the $(100)$, and $x_2$, and $x_3$ axes
are chosen along the $(010)$ and $(001)$ directions,
correspondingly. The Green's function of Eq.~(\ref{eq:1}) is given
by \bea\label{eq:2}
&&\hat G^0_K=\frac{g_++g_-}{2}\hat\sigma_0 + \frac{g_+-g_-}{2|s|}\sum_{i=1,2}{s_i}\hat\sigma_i,\nn\\
&&g_{\pm}=\frac{1}{i\omega_n-\e_{\bk,\pm}},~~\e_{\bk,\pm}=v_F(k-k_F)\pm|s|,\eea
where $|s|=\sqrt{s_1^2+s_2^2}$.

Using the eigenvectors of Hamiltonian (\ref{eq:1})
\beq\label{eq:33} \psi_{\mu}(\vec r)=e^{i\bk\cdot\vec r}\left(
\begin{array}{c}
i\nu e^{-i\phi_\bk}\\1
\end{array}
\right), \eeq where $\tan\phi_\bk=-s_1/s_2$, we obtain for the
matrix elements of Raman scattering
\bea\label{eq:gamma} \gamma_{\mu\mu'}&=&i\delta_{\bk-\bk'-\bq}
\left[n_1\left\{i\nu'e^{-i\phi_{\bk+\bq}}-i\nu
e^{i\phi_{\bk}}\right\}\right.\nonumber\\
&&+ n_2\left\{-\nu'e^{-i\phi_{\bk+\bq}}-\nu
e^{i\phi_{\bk}}\right\}\nonumber\\
&&\left.+ n_3\left\{\nu\nu'
e^{-i(\phi_{\bk+\bq}-\phi_\bk)}-1\right\}
\right],\nn\\
\langle e^{i\bq.\vec r}\rangle_{\mu\mu'}&=&\delta_{\bk-\bk'-\bq}
\left[\nu\nu' e^{-i(\phi_{\bk+\bq}-\phi_\bk)}+1\right]. \eea
We are interested in the small-$q$ limit, when the $q$-dependent
terms in the dispersions of CSWs corrections to the frequencies of
these waves at $q=0$, which implies that $v_Fq\ll \Omega$. At the
same time, both the magnetic field and SOC are assumed to be weak
compared to the Fermi energy (in appropriate units), i.e.,
$\Omega\ll v_Fk_F$. Combining the two inequalities above, we
obtain \beq q\ll \Omega/v_F\ll k_F. \label{ineq} \eeq The first
part of this inequality ensures that the diagonal elements of
$L^0_{\mu\mu'}$ in Eq.~(\ref{eq:Raman1}) are small by charge
conservation: $L^0_{\bk,\pm;\bk+\bq,\pm}\propto q$. Therefore, the
main contribution to the cross-section in this limit comes from
off-diagonal terms with $\nu\nu'=-1$, which correspond to
processes that flip spin/chirality. The second part of
Eq.~(\ref{ineq}) implies that $\phi_{\bk+\bq}$ can be approximated
by $\phi_{k}$, upon which the off-diagonal matrix elements are
simplified to
\begin{subequations}
\bea \gamma_{\mu\mu'}&\approx&-2i\delta_{\bk-\bk'-\bq}\nonumber\\
&&\times\left[in_1\nu\cos\phi_\bk+ +in_2\nu\sin\phi_\bk+ n_3
\right],\\
|\gamma_{\mu\mu'}|^2&\approx& 4\delta_{\bk-\bk'-\bq} \nonumber\\
&&\times\left[(n_1\cos\phi_\bk+n_2\sin\phi_\bk)^2+n_3^2\right],\\
\langle e^{i\bq\cdot\vec
r}\rangle_{\mu\mu'}&\approx&\nu\nu'+1=0.\label{eq:gamm2}\eea
\end{subequations}

By the same argument, the diagonal terms in $Z$ and $\bar Z$ in
Eq.~(\ref{eq:Raman1}) are small as $q$, while the off-diagonal
terms are small because they contain off-diagonal matrix elements
$\langle e^{i\bq\cdot\vec r}\rangle_{\mu\mu'}$, which are small by
Eq.~(\ref{eq:gamm2}). Therefore, the second term in
Eq.~(\ref{eq:Raman1}) can be neglected compared to the first one,
and the cross-section  is
reduced to
\begin{widetext}
\beq\label{eq:th} \frac{d^2\mathcal{A}}{d\Omega
d\mathcal{O}}\propto  \sum_{\mu\mu'}|\gamma_{\mu\mu'}|^2{\rm
Im}L^0_{\mu\mu'} = \int_{\bk}\left[n_1^2\cos^2\phi_\bk +
n_2^2\sin^2\phi_\bk+n_1n_2\sin2\phi_\bk + n_3^2\right]
\int_\omega(g_+\tilde g_-+g_+\tilde g_-), \eeq
\end{widetext}
where $\int_{\bk}\equiv\int{d^2k}/{(2\pi)^2}$;
$\int_{\omega}\equiv d\omega/2\pi$;
$g_{\nu}=1/(i\omega-\e_{\bk,\nu})$ and $\tilde
g_{\nu}=1/(i\{\omega+\Omega\}-\e_{\bk+\bq,\nu})$. The various
terms in the equation above can be expressed via the components of
the spin-spin correlation function in the chiral basis:
\bea\label{eq:tt}
\chi^{0}_{11}(Q)&=&-\int_\bk\int_\omega(g_+\tilde g_-+g_+\tilde g_-)\cos^2\phi_{\bk},\nn\\
\chi^{0}_{22}(Q)&=&-\int_\bk\int_\omega(g_+\tilde g_-+g_+\tilde g_-)\sin^2\phi_{\bk},\nn\\
\chi^{0}_{12}(Q)&=&-\int_\bk\int_\omega(g_+\tilde g_-+g_+\tilde g_-)\sin\phi_{\bk}\cos\phi_{\bk},\nn\\
\chi^{0}_{21}(Q)&=&{\chi^0_{12}(Q)},\nn\\
\chi^{0}_{33}(Q)&=&-\int_\bk\int_\omega(g_+\tilde g_-+g_+\tilde
g_-), \eea Making use of the definitions, we re-write
Eq.~(\ref{eq:th}) as \beq \frac{d^2\mathcal{A}}{d\Omega
d\mathcal{O}}\propto n_1^2\chi^0_{11} +
n_2^2\chi^0_{22}+n_1n_2(\chi^0_{12}+\chi^0_{21}) +
n_3^2\chi^0_{33}, \label{th1} \eeq

Accounting for the exchange part of {\em eei} amounts simply to
replacing the free-electron spin-spin correlation function by the
renormalized ones, i.e., to replacing the bare bubbles in
Fig.~\ref{fig:two} by the shaded ones: \beq
\frac{d^2\mathcal{A}}{d\Omega d\mathcal{O}}\propto n_1^2\chi_{11}
+ n_2^2\chi_{22}+n_1n_2(\chi_{12}+\chi_{21}) + n_3^2\chi_{33}.
\label{th2} \eeq The poles of the renormalized components of the
spin susceptibility tensor correspond to CSWs. This is the main
point of departure from the theory of Ref.~\onlinecite{Magarill},
which accounts for plasmons but not for CSWs. On a technical
level, renormalized $\chi_{ij}$ can be found either from the
equations of motion of the FL theory
\cite{Shekhter,Ali1,kumar:2017} or from Random Phase Approximation
(RPA) in the spin channel. \cite{SM1}
Equation (\ref{th2}) is valid
at finite $q$ provided that $q\ll
k_F$.

Suppose for a moment that SOC is absent while the magnetic field
is applied along the $x_1$-axis. In this case, the $q=0$
susceptibility components satisfy $\chi_{22}=\chi_{33}$  and
$\chi_{11}=\chi_{12}=0$, due to conservation of spin component
along the direction of the field.\cite{SM_ESR} [At small but finite $q$,  $\chi_{11}$ and $\chi_{12}$ are also finite but small (in proportion to $q$) and can be neglected.] The Raman cross-section is then
proportional to $(n_2^2+n_3^2)\chi_{22}$. Because of
spin-rotational symmetry in the absence of the field, $\chi_{22}$
is proportional to the transverse susceptibility
$\chi_\perp=\chi_{+-}+\chi_{-+}$. \cite{Das,perez} The pole of
$\chi_{\perp}$ corresponds to the SL
mode.\cite{silin:1958,leggett:1970,statphysII} The direction of
$\vec n=\vec{\rm e}_{{\rm I}}\times \vec{\rm e}_{{\rm S}}$ affects
only the magnitude of the Raman signal but not its profile.

In the presence of SOC, however, the result for the Raman
cross-section does not reduced to $\chi_\perp$. Instead, the
partial components of $\chi_{lm}$ contribute to the cross-section
with weights determined by the direction of $\vec n$. The poles of
$\chi_{lm}$ now correspond to CSWs shown in Fig.~\ref{fig:1}{\em
a}, one of them being adiabatically connected to the SL mode. In
general, equations of motion for the three components of
magnetization are coupled to each other, and hence all the
components of $\hat \chi$ have poles at the same frequencies but
with different residues. The profile of the spectrum is determined
by the residues as well as by the relative weights of the
geometrical factors $n_in_j$ as given in Eq. (\ref{th2}).

\section{Symmetry arguments for dispersion of chiral spin
waves}\label{sec:symm} In this section, we show that the general
form of the CSW dispersion can be determined solely by symmetry
and dimensional analysys. The role of the microscopic theory,
presented in Sec.~\ref{sec:micro}, amounts then to fixing the
dimensionless functions of the exchange interaction parameter(s),
which enter this form.

As SOC and magnetic field are weak, the interaction of two
quasiparticles with momenta $\bk$ and $\bk'$ can be described by
an $SU(2)$-invariant Landau function \beq \hat F(\vartheta)=\hat
\sigma_0\hat\sigma'_0F^s(\vartheta)+\hat{\vec\sigma}\cdot\hat{\vec\sigma}'F^a(\vartheta),\eeq
where $\vartheta=\theta_\bk-\theta_{\bk'}$. Within this
approximation, the charge and spin sectors of the theory are
decoupled, and we focus on the exchange interaction parameterized
by $F^a(\vartheta)$.

To set the stage, we discuss  the SL mode in the absence of SOC.
Kohn's theorem protects the $q=0$ term in the dispersion from
renormalization by {\em eei};\cite{kohn,yafet} therefore,
$\Omega_0(\theta_{\bB} )=\Delta_Z$  in Eq.~(\ref{res1}).  Now we
consider the dispersion at small but finite $q$: $q\ll
\Delta_Z/v_F$. In the absence of SOC, rotations in the orbital and
spin spaces are independent, which excludes the dot product
$\bq\cdot \bB$ and any power of it thereof. Therefore, the
dispersion starts with a $q^2$ term. Combining the symmetry
arguments with dimensional analysis, we find \bea
\Omega(\bq,\bB)=\Delta_Z+a_2(\{F^a\})\frac{(v_Fq)^2}{\Delta_Z},\label{eq:mt3}
\eea where $a_2$ is a dimensionless function which depends on the
angular harmonics of $F^a(\vartheta)$:
$a_2(\{F^a\})=a_2(F^a_0,F^a_1,F^a_2\dots)$. A precise form of
$a_2$ is determined by the microscopic theory.
\cite{silin:1958,mineev}

We now apply the same reasoning to CSWs. If only Rashba SOC is
present, there is no preferred in-plane direction, hence a
linear-in-$q$ term is absent while the quadratic term is
isotropic.  There are three CSWs in this case, whose dispersions
for $q\ll\Delta_R/v_F$ can be written as \beq
\Omega_\alpha(\bq,\vec 0)=\tilde a_{0}^\alpha(\{F^a\})
{|\Delta_R|}+\tilde a_{2}^\alpha(\{F^a\})\frac{(v_Fq)^2}{
{|\Delta_R|}},\eeq  where $\alpha=1\dots3$ and $\tilde
a^\alpha_{0,2}$ are some other dimensionless functions of the FL
parameters. Since Kohn's theorem does not apply in the presence of
SOC, $\tilde a_{0}^\alpha \neq 1$. Explicit forms of the functions
$\tilde a^\alpha_{0,2}$ were found in Refs.
\onlinecite{Ali1,Zhang,SM1}. If only Dresselhaus SOC is present,
the Hamiltonian can be transformed to the Rashba form by replacing
$\theta_\bk\rightarrow \pi/2-\theta_\bk$. Therefore, the CSWs are
the same in both cases although the Hamiltonians have different
symmetries.

If both Rashba and Dresselhaus types of SOC are present, the $q=0$
term in the dispersion remains isotropic because in the
absence of the magnetic field, the limit of $q=0$ can be reached along any direction in the plane), while a
linear-in-$q$ term is still not allowed by symmetry.
Indeed, since the dispersion must be analytic in $q$, a term of
the type $|\bq|$ is not allowed, and the linear term could only be
of the form $c_1q_1+c_2q_2$ with $c_{1,2}=\text{const}$. However,
such a form does not obey the symmetries of the $D_{2d}$ group
(rotation by $\pi$ and reflection about the diagonal plane). The
quadratic term can have an anisotropic part of the form
$q_1q_2\propto\sin2\theta_{\bq}$, but we already know that such a
term is absent if only Dresselhaus SOC is present. Therefore, such
a term can only arise due to interplay between Rashba and
Dresselhaus types of SOC, and its should be proportional to the
product $\Delta_R\Delta_D$. Hence the anisotropic part is small
compared to the isotropic, $q^2$ part.

Finally, let both types of SOC {\em and} the magnetic field be
present but, in agreement with the experimental conditions of
Refs. \onlinecite{BBX1,BBX2,BBX3}, {we focus on the case when the
Zeeman energy is largest scale in the problem, i.e.,
}$\Delta_Z\gg\Delta_R,\Delta_D$. The $q=0$ term in the dispersion
then depends on the orientation of $\bB$ in the plane; the
$D_{2d}$-symmetry forces this dependence to be of {the} $\sin
2\theta_\bB$ form. Since the anisotropic part of
$\Omega_0(\theta_\bB)$ is non-zero only if both types of SOC are
present, the coefficient of the $\sin 2\theta_\bB$ term must be on
the order of $\Delta_R\Delta_D/\Delta_Z$. In addition to the
anisotropic part, there are also isotropic corrections of order
$\Delta_R^2/\Delta_Z$ and $\Delta_D^2/\Delta_Z$.

To lowest order in SOC, the form of the linear term is determined
by the symmetries of the Rashba (group $C_{\infty v}$) and
Dresselhaus (group $D_{2d}$) types of SOC. In both cases, we need
to form a scalar ($\Omega$) out of a polar vector ($\bq$) and a
pseudovector ($\bB$). In the $C_{\infty v}$ group, this is only
possible by forming the Rashba invariant $B_1q_2-B_2q_1\propto
\sin(\theta_\bq-\theta_\bB)$, which is the same term as in the
original Rashba Hamiltonian with $\hat{\vec\sigma}\to \bB$.
Likewise, the only possible scalar in the $D_{2d}$ group is the
Dresselhaus invariant $B_1q_1-B_2q_2\propto
\cos(\theta_\bq+\theta_\bB)$. The quadratic term in the high-field
limit can be taken the same as the quadratic term in the SL mode,
Eq.~(\ref{eq:mt3}).\cite{FN2}

Combining together all the arguments given above, we arrive at the
following form of the coefficients in Eq.~(\ref{res1}) \bea
\Omega_0(\theta_{\bB})&=&\Delta_Z+a_0(\{F^a\})\frac{\Delta_R^2+\Delta_D^2}{\Delta_Z}\nn\\
&&~~~~~~~~~~+\tilde a_{0}(\{F^a\}) \frac{\Delta_R\Delta_D}{\Delta_Z}\sin2\theta_\bB,\nn\\
w(\theta_\bq,\theta_\bB)&=&v_Fa_1(\{F^a\})\left[\frac{\Delta_R}{\Delta_Z}
\sin(\theta_\bq-\theta_\bB)\right.\nn\\
&&~~~~~~~~~~\left.+\frac{\Delta_D}{\Delta_Z}\cos(\theta_\bq+\theta_\bB)\right],\nn\\
A&=&a_2(\{F^a\})\frac{v_F^2}{\Delta_Z}, \label{conj} \eea where
$a_0$, $\tilde a_0$, $a_1$ and $a_2$ are dimensionless functions.
We will now perform a microscopic calculation to confirm the above
from for the dispersion of collective modes. The linear-in-$q$
term of the dispersion is the most interesting one, as it is the
only term that breaks the symmetry between $\Delta_R$ and
$\Delta_D$. This allows one to extract separately the Rashba and
Dresselhaus components of SOC from the spectrum of the collective
modes. Note that this information cannot be deduced from the $q=0$
term alone.

\section{Microscopic calculation of chiral spin wave
dispersion}\label{sec:micro} In this section, we confirm
Eq.~(\ref{conj}) by an explicit calculation within the $s$-wave
approximation, in which the Landau function contains only the
zeroth harmonic: $F^a(\theta)=F_0^a$. In this case, the FL theory
is equivalent to the Random Phase Approximation (RPA) in the spin
channel,\cite{SM1,SM_ESR} and we adopt the RPA method for
convenience. In this scheme, the shaded bubbles in Fig.
\ref{fig:two} are approximated by the ladder series of
diagrams.\cite{SM1,SM_ESR} Summing up this series, we obtain for
the tensor of the spin susceptibility \bea
\hat\chi(Q)=\frac{(g\mu_B)^2}{4} \hat\chi^*(Q)\left[\hat
I+\frac{F_0^a}{2N_F}\hat\chi^*(Q)\right]^{-1}, \eea where
$Q=(\bq,i\Omega_n)$, $g$ is the effective Land{\'e} factor,
$\mu_B$ is the Bohr magneton, $N_F$ is the density of states at
the Fermi surface, $\hat I$ is the $3\times 3$ identity matrix,
$\chi^*_{ij}(Q)=-\int_K\text{Tr}\left[\hat\sigma_i \hat G^*_K
\hat\sigma_j \hat G^*_{K+Q}\right]$, $K\equiv (\bk,i\omega_m)$ and
$G^*_K$  differs from $G^0_K$ in Eq.~(\ref{eq:2}) only in that the
bare Zeeman energy is replaced by the renormalized one,
$\Delta^*_Z=\Delta_Z/(1+F_0^a)$, while $\Delta_R$ and $\Delta_D$
are not renormalized in the $s$-wave approximation.
\cite{raikh:1999,saraga:2005,Shekhter} The collective modes
correspond to the poles of $\hat\chi$ and can be found as roots of
the equation \beq\label{eq:ins} {\rm Det}\left[\hat
I+\frac{F_0^a}{2N_F}\hat\chi^*(\bq,i\Omega_n\to \Omega+i
0^+)\right]=0. \eeq The details of solving this equation are
purely mathematical in nature and are presented in Appendices \ref{app:bubble}, \ref{app:SL}, and
\ref{app:CSW}. Here, we only state that the results indeed
coincide with those in Eq.~(\ref{conj}) and give explicit
expressions for the dimensionless functions, which read \bea
\tilde a_0(F_0^a)&=&-\frac{(1+F_0^a)(2+F_0^a)}{2F_0^a},\nn\\
a_0(F_0^a)&=&\frac{(1+F_0^a)(2+3F_0^a)}{4F_0^a},\nn\\
a_1(F_0^a)&=&-\frac{(1+F_0^a)^2\left[(4+F_0^a)(1+F_0^a)+(F_0^a)^2\right]}{F_0^a(2+F_0^a)^2},
\nn\\
a_2(F_0^a)&=&\frac{(1+F_0^a)^2}{2F_0^a}. \label{coeff} \eea [The
forms of $a_0$ and $a_2$  coincide with the results of
Refs.~\onlinecite{SM_ESR} and \onlinecite{mineev}, respectively,
in the $s$-wave approximation.] For repulsive {\em eei}, $F_0^a<0$
and the quadratic term in the dispersion is always negative, while
the sign of the interaction-dependent prefactor of the linear term
[$a_1(F^a_0)$] is positive. However, the overall sign of the
linear term depends on the signs of the Rashba and Dresselhaus
couplings, as well as on the orientations of $\bq$ and $\bB$.

The continuum of SFE corresponds to interband transitions and is
given by a set of $\Omega$ that satisfy
$\Omega=|\e_{\bk,+}-\e_{\bk,-}|$ for all states on the FS. Because
$\e_{{\blue \bk},\pm}$ vary around the FS, $\Omega$ varies between
the minimum ($\Omega_{\min}$) and maximum ($\Omega_{\max}$)
values, which determine a finite width of the continuum even at
$q=0$ (see Fig.~\ref{fig:1}a).   In the presence of $\vec B$, both
$\Omega_{{\min}} $ and $\Omega_{{\max}} $ depend on $\theta_\bB$.
The anisotropic part of $\e_{\bk,\pm}$ is given by
$\Delta_R\Delta_D\sin2\theta_\bk$
$+\Delta_R\Delta^*_Z\sin(\theta_\bk-\theta_\bB)$$+\Delta_D\Delta^*_Z\cos(\theta_\bk+\theta_\bB)$.
As one can see, changing $\theta_\bB\rightarrow\theta_\bB+\pi$ can
be compensated by $\theta_\bk\rightarrow\theta_\bk+\pi$, and thus
$\Omega_{\min,\max}(\theta_\bB)$ have a period of $\pi$. Figure
\ref{fig:1}b shows $\Omega_{\min,\max}(\theta_\bB)$ for a range of
the magnetic fields.

\section{Comparison with Experiment}
\label{sec:exp}
We are now in a position to apply our results to recent Raman data
on a Cd$_{1-x}$Mn$_x$Te quantum well.\cite{BBX1,BBX2,BBX3} In these
experiments, $\bq$ and $\bB$ are chosen to be perpendicular to
each other, i.e.,  $\theta_{\bB}-\theta_\bq=\pm\pi/2$.
Accordingly, the dispersion in Eqs.~(\ref{res1}) and (\ref{conj})
is simplified to \bea
\Omega(\bq,\bB)&=&\Delta_Z\left[1+a_0(r^2+d^2)+\tilde a_0 rd \sin
2\theta_\bB\right.\nn
\\&&\left.\pm a_1\left(r-d\sin2\theta_{\vec B}\right)\frac{v_Fq}{\Delta_Z}+a_2\frac{(v_Fq)^2}{\Delta_Z}
\right].
\nn\\
\label{fit} \eea where $r=\Delta_R/\Delta_Z$ and
$d=\Delta_D/\Delta_Z$. Up to the angular dependence of the mode
mass, this form was conjectured in Refs. \onlinecite{BBX2,BBX3} on
phenomenological grounds. The measured frequency of the mode at
$q=0$ gives $\Delta_Z=0.4$\;meV at $B=2$ T. For the number density
of $n=2.7\times10^{11}$cm$^{-2}$ and effective mass of
$m^*=0.1m_e$ ($m_e$ is the bare electron mass),
$k_F=1.3\times10^{-2}$\AA$^{-1}$ and $v_F=1.0\;\text{eV}\cdot$\AA.
The range of $v_Fq$ is from $0$ to  $0.6$ meV.

\begin{figure}[htp]
$\begin{array}{cc}
\includegraphics[width=0.54\columnwidth]{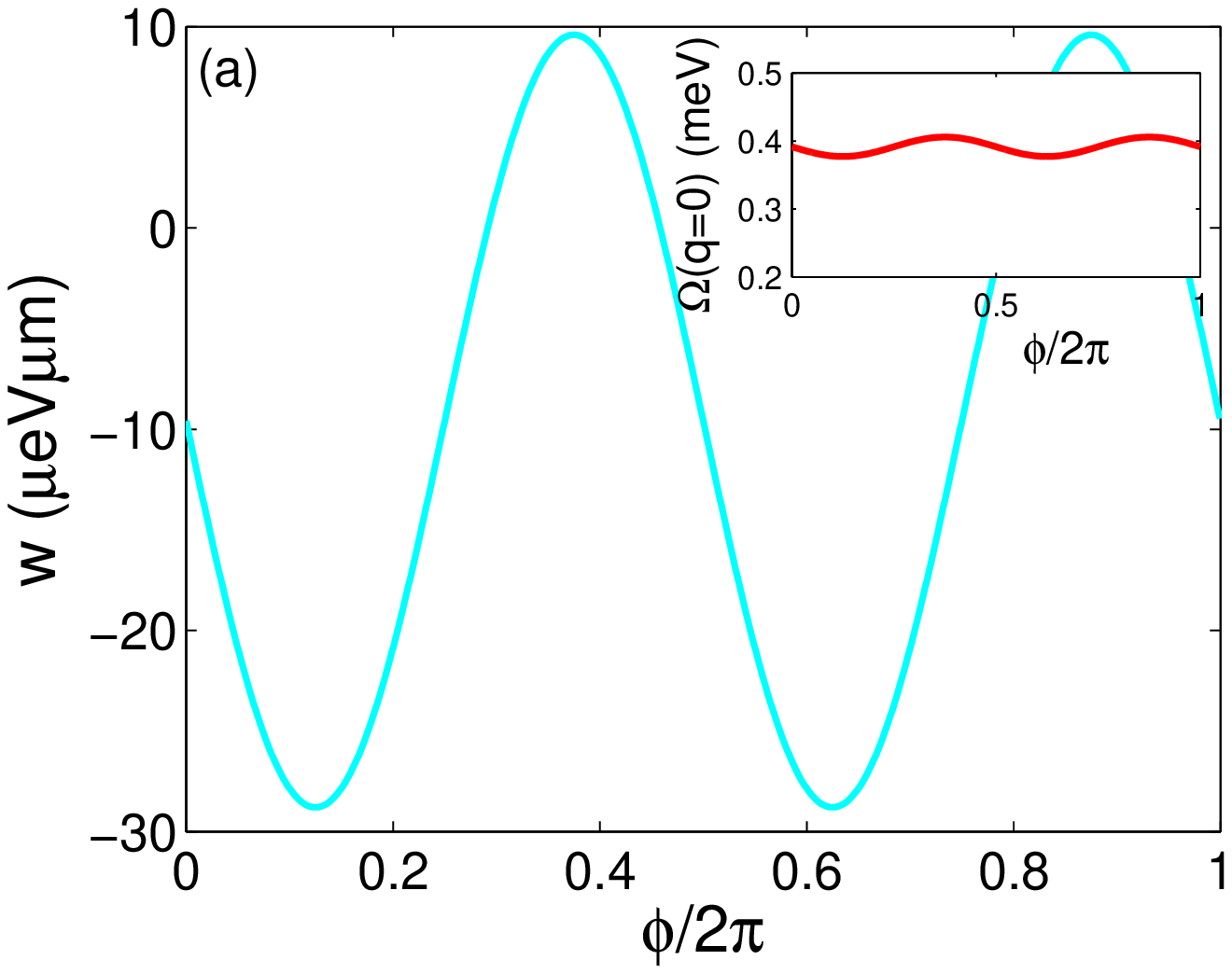}&
\includegraphics[width=0.46\columnwidth,,height=1.4in]{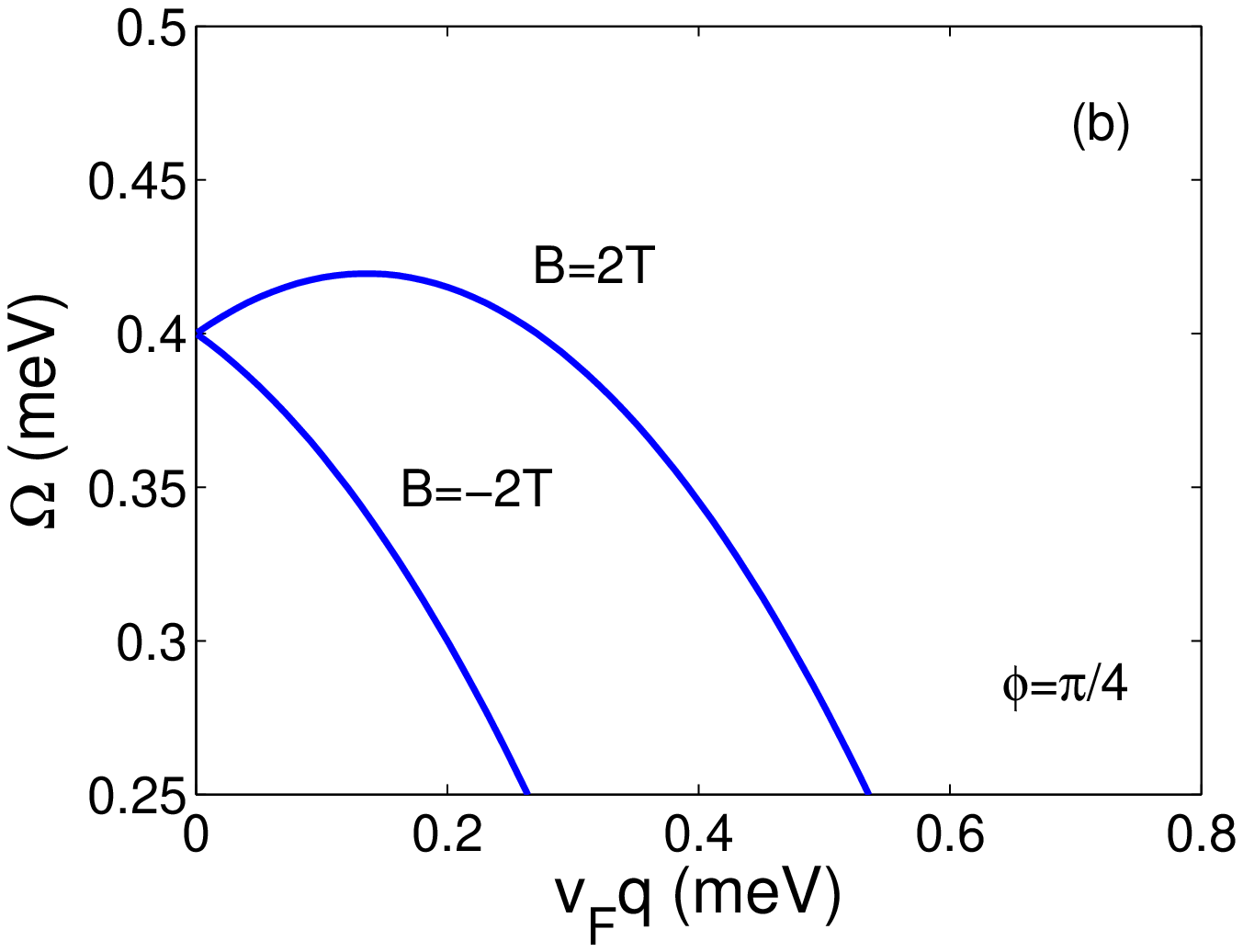}
\end{array}$
\caption{ \label{fig:2} a) Variation of the CSW velocity $w$, as
defined in Eq.~(\ref{res1}), with the angle $
\phi=\theta_{\bB}-\pi/2$. Inset: variation of the frequency at
$q=0$ with $\phi$.  b) Dispersion of a CSW in the CdMnTe quantum
well in the magnetic field of  $2$ T  for $\phi=\pi/4$. To be
consistent with experimental geometry, we fixed $\vec B\perp\bq$.
The negative sign of the field implies flipping its direction.}
\end{figure}

The theoretical results for the mass term, $\Omega_0(\theta_{\vec
B})$, and spin-wave velocity, $w(\pm\pi/2+\theta_\bB,\theta_\bB)$
[cf. Eq.~(\ref{res1})] are plotted in Fig.~\ref{fig:2}{\em a}.
Using the parameters specified above, the best agreement with the
data  is achieved by choosing $F_0^a=-0.41$, $\Delta_R\approx
0.05$ meV, and $\Delta_D\approx 0.1$ meV, which corresponds to the
Rashba and Dresselhaus coupling constants of $\alpha=1.9$
meV$\cdot$\AA\; and $\beta=3.8$ meV$\cdot$\AA. The values of
$\alpha$ and $\beta$ are very close to those found in Ref.
\onlinecite{BBX3}. An estimate for $F^a_0$, obtained using a
screened Coulomb interaction with a dielectric constant
${\epsilon}=10.0$ for a CdTe quantum well,\cite{eps} yields
$F_0^a\approx-0.35$, which is pretty close to the number that
reproduces the experimental data. The calculated $q$-dependence of
the mode (Fig.~\ref{fig:2}b) also reproduces the experimentally
observed one very well.

Based on this agreement between the theory and experiment, we
argue that the collective mode observed in Refs.
\onlinecite{BBX1,BBX2,BBX3} is, in fact, one of the sought-after
CSWs,\cite{Shekhter,Ali1,Zhang,SM1} probed in the regime of a
strong magnetic field. The same type of experiments performed at
lower fields on systems with stronger SOC and/or higher
mobilities, e.g., on {GaAs/GaAlAs} or InGaAs/InAlAs, should reveal
the whole spectrum shown in Fig. \ref{fig:1}a.

Comparison with other theoretical approaches is now in order. The
phenomenological model of Refs. \onlinecite{BBX2} and
\onlinecite{BBX3} describes the data assuming very strong (up to a
factor of $6.5$) renormalization of SOC by many-body effects,
which is not consistent with the moderate ($<2$) values of
parameter $r_s$ in Cd$_{1-x}$Mn$_{x}$Te. Our microscopic theory
explains the data without such an assumption. In addition, it is
argued in Ref.~\onlinecite{BBX2} and \onlinecite{BBX3} the entire
$q$-dependence of a collective-mode spectrum in the presence of
SOC can be obtained by a linear shift of the quadratic term in the
SL dispersion: $Aq^2\to A|\vec q+\vec q_0|^2$. We see from
Eq. (\ref{coeff}) that this would require $|a_1|=2|a_2|$. This is
reproduced only in the weak coupling limit $|F_0^a|\ll1$.

\noindent {\em Note.} When this manuscript was almost completed, a
subset of authors of Refs. \onlinecite{BBX2} and \onlinecite
{BBX3} announced a first-principles study in
Ref.~\onlinecite{BBX4}, in which they also identified the
$\pi$-periodic modulation of the mode frequency at $q=0$ as a
second-order effect in SOC.

\section{Conclusions}\label{sec:con}
In conclusion, we developed a microscopic theory of Raman
scattering in a two-dimensional  Fermi liquid in the presence of
both Rashba and Dresselhaus spin-orbit couplings, and subject to
an in-plane magnetic field. Interplay between an exchange part of
the electron-electron interaction and spin-orbit coupling leads to
resonance peaks at frequencies corresponding to novel collective
modes--chiral spin waves. We derived the polarization dependence
of the Raman signal and showed that the Raman spectrum can be used
to uniquely determine different components of spin-orbit coupling
by measuring a characteristic linear-in-$q$ term in the dispersion
of chiral spin waves. Our theory describes quantitatively all the
features of the Raman signal observed recently on a CdTe quantum
well.\cite{BBX1,BBX2,BBX3} The formalism developed here can be
readily extended to other two-dimensional systems with broken
inversion symmetry, such as graphene on
transition-metal-dichalcogenide substrates and surface states of
topological insulators/superconductors.

\acknowledgements The authors are grateful to Y. Gallais, A.
Kumar, I. Paul, F. Perez, and C. A. Ullrich for stimulating
discussions, and to M. Imran for his help at the initial stage of
this work.
\appendix
\section{Computational details}
The general form of the dispersion a CSW is established in
Eqs.~(\ref{res1}) and (\ref{conj}) using symmetry and dimensional
analysis. Here, we confirm this form by an explicit calculation
which yields the forms of functions $a_0$, $\tilde a_0$, $a_1$,
and $a_2$. We start from Eq.~(\ref{eq:ins}) and find the
expansions of the spin-spin correlation functions,
$\chi^*_{ab}(Q)$, to order $q^2$. Unless stated otherwise, all
frequencies in the Appendix are the Matsubara ones. Whenever it
does not lead to a confusion, the Matsubara  index $n$ will be
suppressed and the frequencies will be denoted simply as
$i\Omega$. Analytic continuation to real frequencies will be done
at the final step. We follow the general scheme developed in
Refs.~\onlinecite{SM1} and \onlinecite{SM_ESR} to find the
dispersions of the collective modes. For brevity, we will switch
to notations where $F_0^a\equiv-u$ and
{$\chi^*_{ab}\equiv-\Pi_{ab}$}.

\subsection{Spin-charge correlation functions in the presence of magnetic field and spin-orbit coupling}
\label{app:bubble} In this Appendix, we consider the general
properties of the spin-charge correlation function defined as \beq
\Pi_{ab}(Q)=\int_{\vec k} \int_\omega \text{Tr} \left[\hat\sigma_a
\hat G^*_{K+Q} \hat\sigma_b \hat G^*_{K}\right], \label{Piab}\eeq
where, as before, $K=(\bk,i\omega_m$), etc., $a,b\in\{0 \dots 3\}$
with the $0^{\rm th}$ component corresponding to the charge and
the $1^{\rm st}\dots 3^{\rm rd}$ components corresponding to three
spin projections. The Green's function $G^*_P$ is obtained from
$G^0_P$ in Eq.~(\ref{eq:2}) by replacing
$\Delta_Z\to\Delta_Z^*=\Delta_Z/(1-u)$. Substituting
Eq.~(\ref{eq:2}) into  Eq.~(\ref{Piab}), we obtain
\begin{widetext} \bea\label{eq:bare}
\Pi_{ab}(Q)&=&\int_{\bk}\int_\omega\left[\frac{g_+\tilde{g}_++g_-\tilde{g}_-}{4}\left\{2\delta_{ab}+
{\rm
Tr}[\hat\sigma_a\hat\sigma_i\hat\sigma_b\hat\sigma_j]\frac{s_i\tilde{s}_j}{|s||\tilde
s|}\right\}+
\frac{g_+\tilde{g}_-+g_-\tilde{g}_+}{4}\left\{2\delta_{ab}-
{\rm Tr}[\hat\sigma_a\hat\sigma_i\hat\sigma_b\hat\sigma_j]\frac{s_i\tilde{s}_j}{|s||\tilde s|}\right\}\right.\nonumber\\
&& + \left. \frac{g_+\tilde{g}_+-g_-\tilde{g}_-}{4}\left\{{\rm
Tr}[\hat\sigma_b\hat\sigma_a\hat\sigma_j]\frac{s_j}{|s|}+ {\rm
Tr}[\hat\sigma_a\hat\sigma_b\hat\sigma_j]\frac{\tilde{s}_j}{|\tilde
s|}\right\}+ \frac{g_+\tilde{g}_--g_-\tilde{g}_+}{4}\left\{{\rm
Tr}[\hat\sigma_b\hat\sigma_a\hat\sigma_j]\frac{s_j}{|s|}-
{\rm Tr}[\hat\sigma_a\hat\sigma_b\hat\sigma_j]\frac{\tilde{s}_j}{|\tilde s|}\right\}\right],\nn\\
\eea\end{widetext} where summation over repeated indices
$i,j\in\{1,2\}$ is implied and a tilde above a quantity means that
its $(2+1)$ momentum is shifted by $Q$ with respect to the
momentum of the corresponding quantity without a tilde, i.e.,
$\tilde g_+\equiv g_+(K+Q)$, etc. Since $s_l$ ($l=1,2$) depend
only on the spatial momentum, it is understood that in this case
$\tilde s_l\equiv s_l(\bk+\bq)$.  Because the magnetic field and
SOC are assumed to be weak, integration over the actual
(anisotropic) FS can be replaced by that over a circular FS of
radius $k_F$. Accordingly, \beq \int_\bk \dots=N_F\int
\frac{d\theta_{\bk}}{2\pi}\int d\xi_\bk\dots\equiv
N_F\int_\theta\int_{\xi_\bk}\dots \eeq where $N_F=k_F/2\pi v_F$ is
the 2D density of states and $\xi_\bk=v_F(k-k_F)$. In the same
approximation, $\e_{\bk,\nu}=\xi_\bk+\nu|s|$, where
$|s|=\sqrt{s_1^2+s_2^2}$ is evaluated at $\xi=0$ but does depend
on the azimuthal angle $\theta_{\bk}$ of $\bk$. Furthermore, we
will be neglecting the difference between $\tilde s_l$ (evaluated
at $\bk+\bq$) and $s_l$ (evaluated at $\bk$): keeping such terms
would amount to higher order corrections. Accordingly, we
approximate $\e_{\bk+\bq,\nu}=\xi_{\bk+\bq}+\nu |\tilde s|\approx
\xi_\bk+\vec v_F \cdot \bq+\nu |s|$, where we also neglected the
${\mathcal O}(q^2)$ term as being higher order in $q/k_F$.

The chiral Green's functions at momentum $\bk+\bq$ need to be
expanded to second order in $\bq$. Within the same approximations
as specified above, \bea\label{eq:gexp} \tilde g_\nu&\equiv&
g_\nu(\bk+\bq,i\omega+i\Omega) \nonumber\\
&=&\bar g_\nu+\partial_j\bar g_\nu q_j +
\frac12\partial_j\partial_{j'}\bar g_\nu q_jq_{j'}
\nonumber\\
&\approx&\bar g_\nu+\bar g_\nu^2\bv_{F}\cdot\bq +\bar
g_\nu^3(\bv_{F}\cdot\bq)^2,
\eea
where $\bar g_\nu\equiv
g_\nu(\bk,i\omega+i\Omega)$ and $\partial_l\equiv
\partial/\partial k_l$.

In what follows, we will need the following integrals
\bea\label{eq:need}
\int_{\xi_\bk} \int_\omega~g_{\nu}\bar g_{\nu'}&=&-\frac{(\nu-\nu')|s|}{i\Omega+(\nu-\nu')|s|},\nn\\
\int_{\xi_\bk} \int_\omega~g_{\nu}\bar g^2_{\nu'}&=&\frac{i\Omega}{(i\Omega+(\nu-\nu')|s|)^2},\nn\\
\int_{\xi_\bk} \int_\omega~g_{\nu}\bar
g^3_{\nu'}&=&\frac{i\Omega}{(i\Omega+(\nu-\nu')|s|)^3}, \eea where
$\int_{{\xi_\bk}}\equiv \int d\xi_{\bk}$. Using these integrals,
we find

\begin{subequations}
\bea \int_{\xi_\bk}\int_\omega g_+\tilde
g_+&=&\frac{\bv_F\cdot\bq}{i\Omega}+
\frac{(\bv_F\cdot\bq)^2}{(i\Omega)^2}
\label{eq:nom3}\\
\int_{\xi_\bk}\int_\omega g_-\tilde g_-&=& \int_{\xi_\bk}\int_\omega g_+\tilde g_+,\label{eq:nom4}\\
\int_{\xi_\bk}\int_\omega g_{\pm}\tilde
g_{\mp}&=&\mp\frac{2|s|}{i\Omega\pm
2|s|}+\frac{i\Omega}{(i\Omega\pm 2|s|)^2}\bv_F\cdot\bq\nonumber\\
&&+ \frac{i\Omega}{(i\Omega\pm
2|s|)^3}(\bv_F\cdot\bq)^2.\label{eq:nom5} \eea
\end{subequations}
Since we put $\tilde s_l = s_l$ in Eq.~(\ref{eq:bare}), the
spin-charge components of the polarization operator, $\Pi_{a0}(Q)$
with $a\neq 0$, are reduced to a combination
$\Pi_{a0}\propto\int_{\xi_{\bk}}\int_{\omega}(g_+\tilde
g_+-g_-\tilde g_-)s_a$, which is equal to zero by
Eq.~(\ref{eq:nom4}). Thus the spin sector is decoupled form the
charge sector in the limit of  $q/k_F\rightarrow 0$. Restricting
to $a,b\in\{1,2,3\}$, we evaluate the traces occurring in
Eq.~(\ref{eq:bare}) as \bea\label{eq:ss}
{\rm Tr}[\hat\sigma_a\hat\sigma_b\hat\sigma_j]s_j&=&i\lambda_{abj} s_j\nn\\
{\rm
Tr}[\hat\sigma_a\hat\sigma_i\hat\sigma_b\hat\sigma_j]s_is_j&=&-2\delta_{ab}
{s}^2 +4s_as_b, \eea where $\lambda_{abc}$ is the Levi-Civita
tensor and it is understood that $s_3=0$. We thus obtain a compact
form of the spin-spin correlation function  with $a,b\in
{1,2,3}$:
\begin{widetext}\beq\label{eq:wtf} \Pi_{ab}(Q)=
{N_F} \int_{\theta_{\bk}}\int_{\xi_\bk}
\int_\omega\left[(g_+\tilde{g}_++g_-\tilde{g}_-)\frac{s_as_b}{|s|^2}+
(g_+\tilde{g}_-+g_-\tilde{g}_+)\left(\delta_{ab}-\frac{s_as_b}{|s|^2}\right)
-i(g_+\tilde{g}_--g_-\tilde{g}_+)\lambda_{abc}\frac{s_c}{|s|}\right],
\eeq where the integrals over $\omega$ and $\xi_\bk$ are to be
substituted from Eqs.~(\ref{eq:nom3}-\ref{eq:nom4}). For further
convenience, we also list explicit formulas for $s_a$ and related
quantities: \bea\label{eq:fulls}
s_1&=&\frac12\left(\Delta_R\sin\theta_{\bk}+\Delta_D\cos\theta_{\bk}-\Delta^*_Z\cos\theta_\bB\right),\nn\\
s_2&=&\frac12\left(-\Delta_R\cos\theta_{\bk}-\Delta_D\sin\theta_{\bk}-\Delta^*_Z\sin\theta_\bB\right),\nn\\
4s_1^2&=&\Delta_R^2\sin^2\theta_{\bk} + \Delta_D^2\cos^2\theta_{\bk}+(\Delta^*_Z)^2\cos^2\theta_\bB + \Delta_R\Delta_D\sin2\theta_{\bk}-2\Delta_R\Delta^*_Z\sin\theta_{\bk}\cos\theta_\bB-2\Delta_D\Delta^*_Z\cos\theta_{\bk}\cos\theta_\bB,\nn\\
4s_2^2&=&\Delta_R^2\cos^2\theta_{\bk} + \Delta_D^2\sin^2\theta_{\bk}+(\Delta^*_Z)^2\sin^2\theta_\bB + \Delta_R\Delta_D\sin2\theta_{\bk}+2\Delta_R\Delta^*_Z\cos\theta_{\bk}\sin\theta_\bB+2\Delta_D\Delta^*_Z\sin\theta_{\bk}\sin\theta_\bB,\nn\\
4{s}^2&=&{4(s_1^2+s^2_2)}=\Delta_R^2+\Delta_D^2+(\Delta^*_Z)^2+2\Delta_R\Delta_D\sin2\theta_{\bk}-2\Delta_R\Delta^*_Z\sin(\theta_{\bk}-\theta_\bB)
-2\Delta_D\Delta^*_Z\cos(\theta_{\bk}+\theta_\bB),\nn\\
4s_1s_2&=&-(\Delta_R^2+\Delta_D^2)\sin\theta_{\bk}\cos\theta_{\bk}
+
(\Delta^*_Z)^2\sin\theta_\bB\cos\theta_\bB-\Delta_R\Delta_D+\Delta_R\Delta^*_Z\cos(\theta_{\bk}+\theta_\bB)+
\Delta_D\Delta^*_Z\sin(\theta_{\bk}-\theta_\bB), \eea where
$\theta_{\bB}$ is the angle between $\vec B$ and the $x_1$ axis.

We can now apply the general result, Eq.~(\ref{eq:wtf}) to
particular situations. In what follows, we will make $v_Fq$ and
$\Omega$ dimensionless by rescaling to
$\Delta^*_Z=\Delta_Z/(1-u)$ and define, for the use in Appendices only,
$r\equiv\Delta_R/\Delta^*_Z$ and $d\equiv\Delta_D/\Delta^*_Z$.
Note that these definitions differ from those in the main text, where
$\Delta_R$ and $\Delta_D$ are rescaled to $\Delta_Z$.

\subsection{Silin-Leggett mode}
\label{app:SL} To test our general formula, we apply it first to a
simple case of the Silin-Leggett mode, the dispersion of which is
known. \cite{mineev} The Silin-Leggett mode is the collective mode
in the absence of SOC ($r=d=0$) and in the presence of $\vec B$.
Although the coordinate system in this case is in fact defined by
the direction of $\vec B$, we choose $\bB$ to point in an
arbitrary direction with respect to a fixed coordinate system.
This will useful for a more general case, when the spin-rotational
symmetry is broken by SOC. In this case, $s=\Delta^*_Z$ is
independent of angle $\theta_{\bB}$. Using Eq.~(\ref{eq:wtf}), we
find:
\bea
\frac{\Pi_{11}(Q)}{-2N_F}&=&\frac{(v_Fq)^2}{2\Omega^2}+\frac{\sin^2\theta_\bB}{\Omega^2+1}
\left(1-\frac{6\Omega^4+3\Omega^2+1}{\Omega^2(\Omega^2+1)^2}\frac{(v_Fq)^2}{2}\right),\label{eq:alpi}\nn\\
\frac{\Pi_{22}(Q)}{-2N_F}&=&\frac{(v_Fq)^2}{2\Omega^2}+\frac{\cos^2\theta_\bB}{\Omega^2+1}
\left(1-\frac{6\Omega^4+3\Omega^2+1}{\Omega^2(\Omega^2+1)^2}\frac{(v_Fq)^2}{2}\right),\nn\\
\frac{\Pi_{33}(Q)}{-2N_F}&=&\frac{1}{\Omega^2+1}
\left(1+\frac{\Omega^6-3\Omega^4}{\Omega^2(\Omega^2+1)^2}\frac{(v_Fq)^2}{2}\right),\nn\\
\frac{\Pi_{12}(Q)}{-2N_F}&=&-\frac{\sin\theta_\bB\cos\theta_\bB}{\Omega^2+1}
\left(1-\frac{6\Omega^4+3\Omega^2+1}{\Omega^2(\Omega^2+1)^2}\frac{(v_Fq)^2}{2}\right),\nn\\
\frac{\Pi_{13}(Q)}{-2N_F}&=&-\frac{\Omega\sin\theta_\bB}{\Omega^2+1}\left(1-\frac{3\Omega^2-1}{(\Omega^2+1)^2}\frac{(v_Fq)^2}{2}\right),\nn\\
\frac{\Pi_{23}(Q)}{-2N_F}&=&\frac{\Omega\cos\theta_\bB}{\Omega^2+1}\left(1-\frac{3\Omega^2+1}{(\Omega^2+1)^2}\frac{(v_Fq)^2}{2}\right).\label{eq:alpi2}
\eea $\Pi_{ab}$ can be written more compactly in a matrix form as
\beq\label{eq:mat2t} \frac{\hat\Pi(Q)}{-2N_F}=
\left(\begin{array}{ccc}
\kappa_a+(\mathcal{B}+\kappa_b)\sin^2\theta_\bB&-(\mathcal{B}+\kappa_b)\sin\theta_\bB\cos\theta_\bB&(\mathcal{C}+\kappa_c)\sin\theta_\bB\\
-(\mathcal{B}+\kappa_b)\sin\theta_\bB\cos\theta_\bB&\kappa_a+(\mathcal{B}+\kappa_b)\cos^2\theta_\bB&-(\mathcal{C}+\kappa_c)\cos\theta_\bB\\
-(\mathcal{C}+\kappa_c)\sin\theta_\bB
&(\mathcal{C}+\kappa_c)\cos\theta_\bB&(\mathcal{B}+\kappa_g)
\end{array}
\right), \eeq \end{widetext}
where \bea \mathcal{B}&=&
\frac{1}{\Omega^2+1}\nn\\
\mathcal{C}&=&-\frac{\Omega}{\Omega^2+1}. \label{bc} \eea The
other definitions are apparent from a term-by-term comparison of
expressions in Eq.~(\ref{eq:alpi}) and the corresponding entries
in Eq.~(\ref{eq:mat2t}). All the $\kappa$'s are corrections that
are small in $v_Fq$. At $v_Fq=0$, the eigenmode equation
Det$(1+u\hat\Pi/2N_F)=0$ has a solution
$\Omega^2+1=s_0\equiv2u-u^2$ such that $\Omega^2=-(1-u)^2$.
Restoring the dimensional frequency and continuing analytically to
real frequencies, we find that  the frequency at the $q=0$ is
simply given by $\Omega=\Delta_Z$, in agreement with Kohn's
theorem. To find corrections to this result at small but finite
$v_Fq$, we look for a solution of the form
$\Omega^2+1=s_0+\kappa_0$. Expanding the eigenmode equation to
linear order in all the $\kappa$'s, we obtain \beq\label{eq:modd}
1-2\mathcal{B}u+(\mathcal{B}^2+\mathcal{C}^2)u^2=
u\left[(1-\mathcal{B}u)(\kappa_a+\kappa_b+\kappa_g)-2u\mathcal{C}\kappa_c\right].
\eeq Solving for $\kappa_0$, we obtain \beq\label{eq:soll}
\kappa_0=\frac{2(1-u)^2}{u}\frac{(v_Fq)^2}{2}. \eeq Restoring the
units and continuing analytically, we obtain the frequency of the
mode at finite $q$ as \beq\label{eq:aaa}
\Omega=\Delta_Z-\frac{(1-u)^2}{2u}\frac{(v_Fq)^2}{\Delta_Z}. \eeq
Relabeling $u\to -F_0^a$, we obtain the coefficient $a_2$, as
given by Eq.~(\ref{coeff}).

\subsection{Chiral spin wave in the high-field limit} \label{app:CSW}
\subsubsection{Frequency of a chiral spin wave at $q=0$}
In this section,
we derive the frequency of the chiral spin wave at $q=0$ and in
the high-field limit, when $\Delta_{R},\Delta_{D}\ll \Delta_Z$  or
$r,d\ll 1$ for dimensionless quantities. [Unless $u$ is very close
$1$, i.e., the system is close to a ferromagnetic transition,
there is no real difference between conditions
$\Delta_{R},\Delta_{D}\ll \Delta_Z$ and $\Delta_{R},\Delta_{D}\ll
\Delta^*_Z$.] With
$\Pi_{ab}(i\Omega)\equiv\Pi_{ab}(i\Omega,\bq=0)$,  we obtain from
Eq.~(\ref{eq:wtf}) \begin{widetext} \bea
\frac{\Pi_{11}(i\Omega)}{-2N_F}&=&\frac{\sin^2\theta_\bB}{\Omega^2+1}
\left[1-3\frac{r^2+d^2}{\Omega^2+1}+2\frac{r^2+d^2-2rd\sin2\theta_\bB}{(\Omega^2+1)^2}\right] + \frac12\frac{r^2+d^2}{\Omega^2+1} + \frac{2rd\sin2\theta_\bB}{(\Omega^2+1)^2},
\nn\\
\frac{\Pi_{22}(i\Omega)}{-2N_F}&=&\frac{\cos^2\theta_\bB}{\Omega^2+1}
\left[1-3\frac{r^2+d^2}{\Omega^2+1}+2\frac{r^2+d^2-2rd\sin2\theta_\bB}{(\Omega^2+1)^2}\right] + \frac12\frac{r^2+d^2}{\Omega^2+1} + \frac{2rd\sin2\theta_\bB}{(\Omega^2+1)^2},\nn\\
\frac{\Pi_{33}(i\Omega)}{-2N_F}&=&\frac{1}{\Omega^2+1}
\left[1-3\frac{r^2+d^2}{\Omega^2+1}+2\frac{r^2+d^2-2rd\sin2\theta_\bB}{(\Omega^2+1)^2}\right] + \frac{r^2+d^2}{\Omega^2+1} + \frac{4rd\sin2\theta_\bB}{(\Omega^2+1)^2},\nn\\
\frac{\Pi_{12}(i\Omega)}{-2N_F}&=&-\frac{\sin\theta_\bB\cos\theta_\bB}{\Omega^2+1}
\left[1-3\frac{r^2+d^2}{\Omega^2+1}+2\frac{r^2+d^2-2rd\sin2\theta_\bB}{(\Omega^2+1)^2}\right] +\frac{rd}{\Omega^2+1} - \frac{2rd}{(\Omega^2+1)^2},\nn\\
\frac{\Pi_{13}(i\Omega)}{-2N_F}&=&-\frac{\Omega\sin\theta_\bB}{\Omega^2+1}
\left[1-2\frac{r^2+d^2}{\Omega^2+1}+2\frac{r^2+d^2-2rd\sin2\theta_\bB}{(\Omega^2+1)^2}\right] -\frac{2rd\Omega\cos\theta_\bB}{(\Omega^2+1)^2},\nn\\
\frac{\Pi_{23}(i\Omega)}{-2N_F}&=&\frac{\Omega\cos\theta_\bB}{\Omega^2+1}
\left[1-2\frac{r^2+d^2}{\Omega^2+1}+2\frac{r^2+d^2-2rd\sin2\theta_\bB}{(\Omega^2+1)^2}\right]
+\frac{2rd\Omega\sin\theta_\bB}{(\Omega^2+1)^2}.\label{eq:expnsn2}
\eea In a matrix form, \beq\label{eq:matr}
{\frac{\hat\Pi(i\Omega)}{-2N_F}}= \left(\begin{array}{ccc}
\kappa_a+(\mathcal{B}+\kappa_b)\sin^2\theta_\bB&\kappa_f-(\mathcal{B}+\kappa_b)\sin\theta_\bB\cos\theta_\bB&(\mathcal{C}+\kappa_c)\sin\theta_\bB + \kappa_d\cos\theta_\bB\\
\kappa_f-(\mathcal{B}+\kappa_b)\sin\theta_\bB\cos\theta_\bB&\kappa_a+(\mathcal{B}+\kappa_b)\cos^2\theta_\bB&-(\mathcal{C}+\kappa_c)\cos\theta_\bB-\kappa_d\sin\theta_\bB\\
-(\mathcal{C}+\kappa_c)\sin\theta_\bB -
\kappa_d\cos\theta_\bB&(\mathcal{C}+\kappa_c)\cos\theta_\bB+\kappa_d\sin\theta_\bB&2\kappa_a+(\mathcal{B}+\kappa_b)
\end{array}
\right), \eeq where $\mathcal{B}$ and $\mathcal{C}$ are the same
as in Eq.~(\ref{bc})  and the $\kappa$'s, which are small in $r$
and $d$,  are again defined by a term-by-term comparison of
Eq.~(\ref{eq:expnsn2}) and the entries in  Eq.~(\ref{eq:matr}). We
seek a solution of the form $\Omega^2+1=s_0 + \kappa_0$, where
$\kappa_0$ is also small in $r$ and $d$. To linear order in
$\kappa$'s, the eigenmode equation reads \beq\label{eq:mod}
1-2\mathcal{B}u+(\mathcal{B}^2+\mathcal{C}^2)u^2=
u\left[(1-\mathcal{B}u)(2\kappa_b+3\kappa_a-\kappa_f\sin2\theta_\bB)
-2u\mathcal{C}(\kappa_c+\kappa_d\sin2\theta_\bB)\right]. \eeq
Solving for $\kappa_0$ we get: \beq\label{eq:sol}
\kappa_0=(r^2+d^2)\left\{\frac{(1-u)(2-3u)}{2u}\right\}-
rd\sin2\theta_\bB{\left\{\frac{(1-u)(2-u)}{u}\right\}}. \eeq
Restoring the units and continuing analytically to real
frequencies, we obtain the coefficients $a_0$ and $\tilde a_0$ in
Eq.~(\ref{coeff}).

\subsubsection{Linear-in-$q$ term in the dispersion}
Now we are interested in all terms to linear order in $r$, $d$,
and $v_Fq$. Accordingly, we need to expand the quantities in
Eq.~(\ref{eq:fulls}) to linear order in these variables:
\bea\label{eq:kkk2}
2s_1&=&\Delta^*_Z\left(-\cos\theta_\bB+r\sin\theta_{{\bk}}+d\cos\theta_{\bk}\right),\nn\\
2s_2&=&\Delta^*_Z\left(-\sin\theta_\bB-r\cos\theta_{{\bk}}-d\sin\theta_{\bk}\right),\nn\\
4s^2_1&=&
(\Delta^*_Z)^2\left(\cos^2\theta_\bB-2r\sin\theta\theta_{{\bk}}\cos\theta_\bB-2d\cos\theta_{ {\bk}}\cos\theta_\bB\right),\nn\\
4s^2_2&=&
(\Delta^*_Z)^2\left(\sin^2\theta_\bB+2r\cos\theta_{{\bk}}\sin\theta_\bB+2d\sin\theta_{{\bk}}\sin\theta_\bB\right),\nn\\
4s^2&=&(\Delta^*_Z)^2\left(1-2r\sin(\theta_{{\bk}}-\theta_\bB)-2d\cos(\theta_{{\bk}}+\theta_\bB)\right),\nn\\
4s_1s_2&=&
(\Delta^*_Z)^2\left[\sin\theta_\bB\cos\theta_\bB+r\cos(\theta_{{\bk}}+\theta_\bB)+d\sin(\theta_{{\bk}}-\theta_\bB)\right].
\eea Furthermore,
 \bea
\bv_F\cdot\bq&=&v_Fq\cos(\theta_{{\bk}}-\theta_{\bq})\nn\\
\int_{\xi_\bk}\int_\omega(g_+\tilde{g}_++g_-\tilde{g}_-)&=&
\frac{2v_Fq}{i\Omega}\cos(\theta_{{\bk}}-\theta_{\bq}),\nn\\
\int_{\xi_\bk}\int_\omega(g_+\tilde{g}_-+g_-\tilde{g}_+)&=&
-\frac{2}{\Omega^2+1}\left[1-\frac{2\tilde
r_{{\theta_\bk}}\Omega^2}{\Omega^2+1}\right]
-\frac{2i\Omega v_Fq}{(\Omega^2+1)^2}\left[\Omega^2-1+\frac{2\tilde r_{{\theta_\bk}}(3\Omega^2-1)}{\Omega^2+1}\right]\cos(\theta_{{\bk}}-\theta_{\bq}),\nn\\
\int_{\xi_\bk}\int_\omega(g_+\tilde{g}_--g_-\tilde{g}_+)&=&
\frac{2i\Omega}{\Omega^2+1}\left[1-\frac{\tilde
r_{{\theta_\bk}}(\Omega^2-1)}{\Omega^2+1}\right]
+\frac{4\Omega^2 v_Fq}{(\Omega^2+1)^2}\left[1-\frac{\tilde r_{{\theta_\bk}}(\Omega^2-3)}{\Omega^2+1}\right]\cos(\theta_{{\bk}}-\theta_{\bq}),\nn\\
\eea where $\tilde r_\vartheta\equiv
r\sin(\vartheta-\theta_\bB)+d\cos(\vartheta+\theta_\bB)$. Let's
further introduce \bea\label{eq:trump}
r_{1,\vartheta}&\equiv&r\sin\vartheta+d\cos\vartheta,\nn\\
r_{2,\vartheta}&\equiv&r\cos\vartheta+d\sin\vartheta,\nn\\
r_{3,\vartheta}&\equiv&
r_{1,\vartheta}\sin\theta_\bB+r_{2,\vartheta}\cos\theta_\bB \eea
such that
$r_{1,\vartheta}\cos\theta_\bB-r_{2,\vartheta}\sin\theta_\bB=\tilde
r_{\vartheta}$. Note that {the pairs} $(\tilde r_\vartheta,
r_{3,\vartheta})$ and $(r_{1,\vartheta},r_{2,\vartheta})$ are
related by a $\theta_\bB$ rotation. This leads to: \bea
\frac{\Pi_{11}(Q)}{-2N_F}&=&\frac{\sin^2\theta_\bB}{\Omega^2+1}
\left[1+\frac{i\Omega(3\Omega^2-1)}{(\Omega^2+1)^2}v_Fq\tilde
r_{\theta_\bq}\right]+
\frac{v_Fq}{i\Omega}\frac{\Omega^2(3\Omega^2+1)}{2(\Omega^2+1)^2}r_{3,\theta_\bq}\sin2\theta_\bB,\label{eq:alpit11}\\
\frac{\Pi_{22}(Q)}{-2N_F}&=&\frac{\cos^2\theta_\bB}{\Omega^2+1}
\left[1+\frac{i\Omega(3\Omega^2-1)}{(\Omega^2+1)^2}v_Fq\tilde
r_{\theta_\bq}\right]-
\frac{v_Fq}{i\Omega}\frac{\Omega^2(3\Omega^2+1)}{2(\Omega^2+1)^2}r_{3,\theta_\bq}\sin2\theta_\bB,\\
\frac{\Pi_{33}(Q)}{-2N_F}&=&\frac{1}{\Omega^2+1}
\left[1+\frac{i\Omega(3\Omega^2-1)}{(\Omega^2+1)^2}v_Fq\tilde r_{\theta_\bq}\right],\\
\frac{\Pi_{12}(Q)}{-2N_F}&=&-\frac{\sin\theta_\bB\cos\theta_\bB}{\Omega^2+1}
\left[1+\frac{i\Omega(3\Omega^2-1)}{(\Omega^2+1)^2}v_Fq\tilde
r_{\theta_\bq}\right]-
\frac{v_Fq}{i\Omega}\frac{\Omega^2(3\Omega^2+1)}{2(\Omega^2+1)^2}
r_{3,\theta_\bq}\cos2\theta_\bB,\\
\frac{\Pi_{13}(Q)}{-2N_F}&=&-\frac{\Omega\sin\theta_\bB}{\Omega^2+1}
\left[1+\frac{2i\Omega(\Omega^2-1)}{(\Omega^2+1)^2}v_Fq\tilde
r_{\theta_\bq}\right]-
\frac{v_Fq}{i\Omega}\frac{\Omega^3r_{3,\theta_\bq}\cos\theta_\bB}{(\Omega^2+1)^2},\\
\frac{\Pi_{23}(Q)}{-2N_F}&=&\frac{\Omega\cos\theta_\bB}{\Omega^2+1}
\left[1+\frac{2i\Omega(\Omega^2-1)}{(\Omega^2+1)^2}v_Fq\tilde
r_{\theta_\bq}\right]-
\frac{v_Fq}{i\Omega}\frac{\Omega^3r_{3,\theta_\bq}\sin\theta_\bB}{(\Omega^2+1)^2}.\label{eq:alpit2}
\eea In a matrix form, \beq\label{eq:mautt}
\frac{\hat\Pi(Q)}{-2N_F}= \left(\begin{array}{ccc}
\kappa_a\sin2\theta_\bB+(\mathcal{B}+\kappa_b)\sin^2\theta_\bB&-\kappa_a\cos2\theta_\bB-(\mathcal{B}+\kappa_b)\sin\theta_\bB\cos\theta_\bB&-(\mathcal{C}+\kappa_c)\sin\theta_\bB-\kappa_d \cos\theta_\bB\\
-\kappa_a\cos2\theta_\bB-(\mathcal{B}+\kappa_b)\sin\theta_\bB\cos\theta_\bB&-\kappa_a\sin2\theta_\bB+(\mathcal{B}+\kappa_b)\cos^2\theta_\bB&(\mathcal{C}+\kappa_c)\cos\theta_\bB-\kappa_d\sin\theta_\bB\\
(\mathcal{C}+\kappa_c)\sin\theta_\bB+\kappa_d\cos\theta_\bB
&-(\mathcal{C}+\kappa_c)\cos\theta_\bB+\kappa_d\sin\theta_\bB&(\mathcal{B}+\kappa_b)
\end{array}
\right). \eeq {Once again $\mathcal{B}$ and $\mathcal{C}$ are the
same as before and the $\kappa$'s are {obtained by a term-term
comparison of} Eqs. (\ref{eq:alpit11}-\ref{eq:alpit2}) and
Eq.~(\ref{eq:mautt})}.

We look for a solution of the form $\Omega^2+1=s_0+\kappa_0$. To
linear order in $\kappa$'s, the eigenmode equation is now of the
{following} form \beq\label{eq:modd}
1-2\mathcal{B}u+(\mathcal{B}^2+\mathcal{C}^2)u^2=2u\left\{(1-\mathcal{B}u)\kappa_b-
u\mathcal{C} \kappa_c\right\} \eeq Solving for $\kappa_0$, we get
\bea\label{eq:soll}
\kappa_0&=&-\frac{2(1-u)^2[(4-u)(1-u)+u^2]}{u(2-u)^2}v_Fq\tilde r_{\theta_\bq}\nonumber\\
&=&-\frac{2(1-u)^2[(4-u)(1-u)+u^2]}{u(2-u)^2}v_Fq
\left(r\sin(\theta_{\bq}-\theta_\bB)+d\cos(\theta_{\bq}+\theta_\bB)\right).
\eea From here, one can read the coefficient $a_1$ as given by
Eq.~(\ref{coeff}).
\end{widetext}

\bibliographystyle{apsrev4-1}

\begin{thebibliography}{10}
\bibitem{intro1} T. P. Devereaux and R. Hackl,``Inelastic light scattering from correlated electrons'',  Rev. Mod. Phys. {\bf 79}, 175 (2007).
\bibitem{intro2} Y. Kajiwara, K. Harii, S. Takahashi, J. Ohe, K. Uchida, M. Mizuguchi, H. Umezawa, H. Kawai, K. Ando, K. Takanashi, S. Maekawa, and E. Saitoh, ``Transmission of electrical signals by spin-wave interconversion in a magnetic insulator'', Nature \textbf{464}, 262 (2010).
\bibitem{raman_book} W. Hayes and R. Loudon, Scattering of Light by Crystals (John Wiley and Sons, 1978).
\bibitem{Rashba} E. I. Rashba and V. I. Sheka, ``Symmetry of Energy Bands in Crystals of Wurtzite Type II. Symmetry of Bands with Spin-Orbit Interaction Included'',
Sov. Phys. Solid State {\bf 3}, 1257 (1961); Yu. Bychkov and E. I. Rashba, ``Properties of a 2D electron gas with lifted spectral degeneracy'', JETP Lett. {\bf 39}, 79 (1984).
\bibitem{Dresselhaus} G. Dresselhaus, ``Spin-Orbit Coupling Effects in Zinc Blende Structures'', Phys. Rev. {\bf 100}, 580 (1955).
\bibitem{Magarill} A.V. Chaplik, L.I. Magarill and R.Z. Vitlina, ``Inelastic light scattering by 2D electron system with SO interaction'', Nanoscale Research Letters \textbf{7}, 537 (2012).
\bibitem{Shekhter} A. Shekhter, M. Khodas, and A. M. Finkelstein, ``Chiral spin resonance and spin-Hall conductivity in the presence of the electron-electron interactions'', Phys. Rev. B \textbf{71}, 165329 (2005).
\bibitem{Ali1}A. Ashrafi and D. L. Maslov, ``Chiral Spin Waves in Fermi Liquids with Spin-Orbit Coupling'', Phys. Rev. Lett. \textbf{109}, 227201 (2012).
\bibitem{Zhang} S.-S. Zhang, X.-L. Yu, J. Ye, and W.-Mi. Liu, ``Collective modes of spin-orbit-coupled Fermi gases in the repulsive regime'', \pra \textbf{87}, 063623 (2013).
\bibitem{SM1}S. Maiti, V. A. Zyuzin, and D. L. Maslov, ``Collective modes in two- and three-dimensional electron systems with Rashba spin-orbit coupling'', Phys. Rev. B \textbf{91}, 035106 (2015).
\bibitem{SM_damping} S. Maiti and D. L. Maslov, ``Intrinsic Damping of Collective Spin Modes in a Two-Dimensional Fermi Liquid with Spin-Orbit Coupling'', \prl {\bf 114}, 156803 (2015).
\bibitem{SM_ESR} S. Maiti, M. Imran, and D. L. Maslov, ``Electron spin resonance in a two-dimensional Fermi Liquid with spin-orbit coupling'', Phys. Rev. B \textbf{93}, 045134 (2016).
\bibitem{kumar:2017} A. Kumar and D. L. Maslov, ``Effective lattice model for collective modes in a Fermi liquid with spin-orbit coupling'',  arXiv:1701.02781.
\bibitem{BBX1} F. Baboux, F. Perez, C. A. Ullrich, I. D'Amico, G. Karczewski, and T. Wojtowicz, ``Coulomb-driven organization and enhancement of spin-orbit fields in collective spin excitations'', Phys. Rev. B \textbf{87}, 121303(R) (2013).
\bibitem{BBX2} F. Baboux, F. Perez, C. A. Ullrich, G. Karczewski, and T. Wojtowicz, ``Electron density magnification of the collective spin-orbit field in quantum wells'', Phys. Rev. B \textbf{92}, 125307 (2015).
\bibitem{BBX3} F. Perez, F. Baboux, C. A. Ullrich, I. D’Amico, G. Vignale, G. Karczewski, and T. Wojtowicz, ``Spin-Orbit Twisted Spin Waves: Group Velocity Control'', Phys. Rev. Lett. \textbf{117}, 137204, (2016).
\bibitem{silin:1958}V. P. Silin, ``Oscillations of a Fermi liquid in a magnetic field'', Sov. Phys. JETP {\bf 6}, 945 (1958).
\bibitem{leggett:1970}A. J. Leggett, ``Spin diffusion and spin echoes in liquid $^3$He at low temperature'', J. Phys. C: Solid State Phys. {\bf 3}, 448 (1970).
\bibitem{statphysII} E. M. Lifshitz and L. P. Pitaevskii, Statistical Physics Part II (Pergamon Press, New York, 1980).
includes only the states within either the conduction or valence band.
\bibitem{Das} S.D. Sarma and D.W. Wang, ``Resonant Raman Scattering by Elementary Electronic Excitations in Semiconductor Structures'', Phys. Rev. Lett. \textbf{83}, 816 (1999).
\bibitem{perez} F. Perez, ``Spin-polarized two-dimensional electron gas embedded in a semimagnetic quantum well: Ground state, spin responses, spin excitations, and Raman spectrum'', Phys. Rev. B \textbf{79}, 045306 (2009).
\bibitem{kohn} W. Kohn, ``Cyclotron Resonance and de Haas-van Alphen Oscillations of an Interacting Electron Gas'', Phys. Rev. \textbf{123}, 1242 (1961).
\bibitem{yafet} Y. Yafet, ``$g$-factors and spin-lattice relaxation'', in Solid State Physics, Vol. 14, edited by F. Seitz and D. Turnbull (Academic, New York, 1963), p. 92.
\bibitem{mineev} See V. P. Mineev, ``Transverse spin dynamics in a spin-polarized Fermi liquid'', \prb {\bf 69}, 144429 (2004), and references therein.
\bibitem{FN2} Strictly speaking, the coefficients of the linear invariants may depend on the direction of $\bB$. This effect, however, occurs only to order $\Delta_R\Delta_D$ and can thus be neglected at small $q$. By the same token, the correction to the $q^2$ term in SL mode is on the order $\Delta_R\Delta_D/\Delta_Z^2$ and can also be neglected.
\bibitem{raikh:1999} G.-H. Chen and M. E. Raikh, ``Exchange-induced enhancement of spin-orbit coupling in two-dimensional electronic systems'', \prb \textbf{60}, 4826 (1999).
\bibitem{saraga:2005}D. S. Saraga and D. Loss, ``Fermi liquid parameters in two dimensions with spin-orbit interaction'', \prb\textbf{72}, 195319 (2005).
\bibitem{eps} I. Strzalkowski, S. Joshi, and C. R. Crowell, ``Dielectric constant and its temperature dependence for GaAs, CdTe, and ZnSe'', Appl. Phys. Lett. \textbf{28}, 350
(1976).
\bibitem{BBX4} S. Karimi, F. Baboux, F. Perez, C.A. Ullrich, G. Karczewski, T. Wojtowicz, ``Spin precession and spin waves in a chiral electron gas: beyond Larmor's theorem'', arXiv:1612.04314 (2016).
\end{thebibliography}

\end{document}